\documentclass[paper]{JHEP}
\usepackage{epsfig}
\usepackage{multirow,booktabs}
\usepackage{amssymb}
\def\beq{\begin{equation}}
\def\beqn{\begin{eqnarray}}
\def\eeq{\end{equation}}
\def\eeqn{\end{eqnarray}}
\def\abs#1{\left|#1\right|}

\def\VEV#1{\left\langle #1\right\rangle}
\def\HW{{\small HERWIG}}
\def\HWpp{{\small HERWIG++}}
\def\PY{{\small PYTHIA}}
\def\NLHW{{\small MC@NLO/HW}}
\def\NLPY{{\small MC@NLO/PY}}

\newcommand\sss{\scriptscriptstyle}
\newcommand\mydot{\!\cdot\!}

\newcommand\as{\alpha_{\sss S}}

\newcommand\logarg{{\cal Q}}

\newcommand\clH{{\mathbb H}}
\newcommand\clS{{\mathbb S}}

\newcommand\EVprjmapp{{\cal P}_{\clH\to\clS}^{(+)}}
\newcommand\EVprjmapm{{\cal P}_{\clH\to\clS}^{(-)}}
\newcommand\EVprjmappm{{\cal P}_{\clH\to\clS}^{(\pm)}}

\newcommand\IEVprjmapp{{\cal P}_{\clS\to\clH}^{(+)}}
\newcommand\IEVprjmapm{{\cal P}_{\clS\to\clH}^{(-)}}
\newcommand\IEVprjmappm{{\cal P}_{\clS\to\clH}^{(\pm)}}
\newcommand\Sfun{{\cal S}}

\newcommand\Sfunp{\Sfun_{+}}
\newcommand\Sfunm{\Sfun_{-}}
\newcommand\Sfunpm{\Sfun_{\pm}}
\newcommand\SfunL{\Sfun_{L}}
\newcommand\stepf{\Theta}
\newcommand\mborn{{\cal M}^{(b)}}

\newcommand\xMCB{\Big|_{\sss {\rm MC}}}
\newcommand\evBB{{\Bigg|_{\rm ev}}}
\newcommand\cntBB{\Bigg|_{\rm ct}}
\newcommand\xMCBB{{\Bigg|_{\sss {\rm MC}}}}

\newcommand\Kone{{\bf 1}}
\newcommand\Kqone{{\bf \tilde{1}}}
\newcommand\Ktwo{{\bf 2}}
\newcommand\Ione{{\cal I}_1}
\newcommand\Iqone{{\cal I}_{\tilde {1}}}
\newcommand\xBpyo{\zeta_1}
\newcommand\xBpyt{\zeta_2}
\newcommand\xBpyi{\zeta_i}
\newcommand\hxBpyo{z_1}
\newcommand\hxBpyt{z_2}

\newcommand\mV{m_V}
\newcommand\pt{p_{\sss T}}
\newcommand\kt{k_{\sss T}}
\newcommand\mt{m_{\sss T}}
\newcommand\muF{\mu_{\sss F}}
\newcommand\muR{\mu_{\sss R}}
\newcommand\xosp{x_{1+}^{(s)}}
\newcommand\xtsp{x_{2+}^{(s)}}
\newcommand\xocpp{x_{1+}^{(c+)}}
\newcommand\xtcpp{x_{2+}^{(c+)}}
\newcommand\xosm{x_{1-}^{(s)}}
\newcommand\xtsm{x_{2-}^{(s)}}
\newcommand\xocmm{x_{1-}^{(c-)}}
\newcommand\xtcmm{x_{2-}^{(c-)}}
\newcommand\xospm{x_{1\pm}^{(s)}}
\newcommand\xtspm{x_{2\pm}^{(s)}}
\newcommand\Hone{(H_1)}
\newcommand\Htwo{(H_2)}
\newcommand\bSigma{\overline{\Sigma}}
\newcommand\hSigma{\widehat{\Sigma}}

\preprint{
 CERN-TH/2010-030
 }
\title{Matching NLO QCD computations with PYTHIA using MC@NLO}
\author{Paolo Torrielli\\
  ITPP, EPFL, CH-1015 Lausanne, Switzerland\\
  E-mail: \email{Paolo.Torrielli@epfl.ch}}
\author{Stefano Frixione%
  \thanks{On leave of absence from INFN, Sez. di Genova, Italy.}\\
  PH Department, TH Unit, CERN, CH-1211 Geneva 23, Switzerland\\
  ITPP, EPFL, CH-1015 Lausanne, Switzerland\\
  E-mail: \email{Stefano.Frixione@cern.ch}}
\abstract{We present the matching between a next-to-leading order computation 
in QCD and the PYTHIA parton shower Monte Carlo, according to the MC@NLO 
formalism. We study the case of initial-state radiation, and consider
in particular single vector boson hadroproduction.
}

\begin{document}

\section{Introduction\label{sec:intro}}
Perturbation theory offers a systematic way to improve theoretical predictions 
for any given infrared-safe observable. Depending on whether the expansion
parameter in the series is $\as$, or is $\as$ times a logarithm or a 
logarithm squared of some function of the kinematics of the hard process,
one obtains a fixed-order result or a resummed result, respectively.
Cross sections expanded up to a certain order in $\as$ and those resummed
are relevant to complementary kinematic regions of the phase-space.
It is therefore convenient to combine the features of these two expansions,
by defining a matched cross section which is equal to the former or to
the latter in the appropriate phase-space region.

Fixed-order results are now generally available at the next-to-leading
order (NLO), which corresponds to including in the cross section 
the coefficients of terms of order $\as^b$ (the leading order or Born level)
and of order $\as^{b+1}$. Thanks to the recent and rapid progress in the
automated treatment of both the real and the virtual contributions
to NLO computations, it is realistic to assume that phenomenological
results at this accuracy will become available in the next few
years for all of the reactions of interest to the LHC physics
programme. The situation is much less encouraging for fixed-order
cross sections at ${\cal O}(\as^{b+2})$ (or NNLO), where only a handful 
of results are available, for very small final-state multiplicities.

As far as resummed results are concerned, it is in general understood
how to achieve a next-to-leading logarithmic (NLL) accuracy,
which is equivalent to including terms proportional to $\as^n\log^{kn}\logarg$
and to $\as^n\log^{kn-1}\logarg$, with $k=1,2$ depending on the nature
of $\logarg$; some results are also known to next-to-next-to-leading
logarithmic accuracy (NNLL). Unfortunately, resummed computations are
in general technically complicated, observable-dependent, and error-prone;
although for some observable classes a semi-automated algorithm
(CAESAR~\cite{Banfi:2004yd}) is available, the overall situation 
is far less satisfactory
than for NLO computations. For this reason, a very appealing alternative
is that provided by Parton Shower Monte Carlos (PSMCs). Although
formally PSMCs are equivalent to a LL-accurate resummation (which may
become NLL in some corners of the phase space, for some PSMCs and
subject to certain restrictions), in practice they are known to do much 
better than that, as comparisons with data and with analytically-resummed
results clearly show. Furthermore, PSMCs are fully flexible, give one the
possibility of including hadronization models in a consistent manner,
and are the workhorses of experimental collaborations, thanks to their
capabilities of simulating fully-realistic final-state configurations
that can undergo detector simulations.

As is well known, the approximations that form the core of PSMCs severely
limit their predictive power in those phase-space regions (corresponding
to multi-jet configurations) which are of interest for most of new-physics
searches at colliders. These limitations can be alleviated by viewing
PSMC predictions as resummed results, to be included with the NLO 
corrections to the relevant production processes into a matched
cross section. The definition of a formalism for matching NLO computations
and MC simulations has attracted a considerable amount of attention.
There are now several proposals, but only two of them, namely 
MC@NLO~\cite{Frixione:2002ik} or POWHEG~\cite{Nason:2004rx},
have made it to the stage of actually implementing several 
hadroproduction processes, and of being routinely used by
experimental collaborations.

The MC@NLO formalism requires the computation of the cross section
predicted by the PSMC at ${\cal O}(\as^{b+1})$. Because of the structure
of PSMCs, the non-trivial information of this computation is actually
process-independent, and is contained in the definitions of the
parton branchings; the process-dependent part is entirely factorized
in the Born matrix element. Thus, one essentially has to perform one
set of computations (since typically initial- and final-state branchings
are treated differently by the Monte Carlos) per PSMC, in order to be
able to match an NLO computation with a parton shower simulation according
to the MC@NLO approach. So far, these computations have been carried
out for the case of \HW\ 
(see refs.~\cite{Frixione:2002ik,Frixione:2003ei,Frixione:2005vw})
and, more recently, for \HWpp~\cite{HWpp}.
In this paper, we present the first results relevant to the matching 
with \PY~6.4. We limit ourselves to considering the case of initial-state
branchings, and present some sample results for the Drell-Yan process.

This paper is organized as follows. In sect.~\ref{sec:mcatnlo} we describe
the various steps necessary for an MC@NLO matching with \PY, from the
definition of the underlying parton-level NLO cross section to the
MC@NLO short-distance cross sections used for the generation of
hard-subprocess events, to be showered by \PY. We present sample
results in sect.~\ref{sec:res}, and we give our conclusions in
sect.~\ref{sec:concl}.

\section{MC@NLO\label{sec:mcatnlo}}
\subsection{NLO parton-level cross section}
The starting point for the construction of MC@NLO is that of writing
the short-distance parton-level cross section according to the subtraction
formalism of refs.~\cite{Frixione:1995ms,Frixione:1997np} (which we shall
call FKS subtraction henceforth). The default procedure in FKS is that
of treating simultaneously (i.e. in one contribution to the
partonic cross section) the two initial-state collinear singularities, 
due to one given final-state parton being collinear to the initial-state 
parton coming from the left or from the right. On the other hand, in
FKS one can also treat these two singularities independently, by 
defining two separate contributions to the short-distance cross section,
each of which corresponds to one of the initial-state singularities;
this procedure is explained in detail in ref.~\cite{Frederix:2009yq}.
As we shall discuss in the following, when an NLO computation
is matched to \PY\ according to the MC@NLO formalism, it turns out
to be convenient to adopt the latter strategy (at variance with the case 
of \HW, where the simultaneous subtraction suffices).

We shall assign the momenta entering the (real-emission) 
partonic subprocesses as follows:
\beq
a(p_1)+b(p_2)\;\longrightarrow\;V(k_1)+c(k_2)\,,
\label{Rproc}
\eeq
with $V$ representing a $W$ or a $Z$ boson, and $a$, $b$, and $c$ being
QCD partons; Born-like processes are simply obtained from eq.~(\ref{Rproc})
by removing $c(k_2)$ from the r.h.s.. Only soft and initial-state collinear 
singularities are present in the processes of eq.~(\ref{Rproc}). Hence,
the $\Sfun$ functions required for a separate treatment of the collinear 
singularities are two and shall be denoted by
\beq
\Sfunp\,,\;\;\;\;\;\;\Sfunm\,,
\eeq
with the following properties: 
\beqn
\Sfunp+\Sfunm&=&1\,,
\label{Sfunnorm}\\
\lim_{\vec{k}_2\parallel\vec{p}_1}\Sfunp\,&=&1\,,
\label{Sfuncolp}\\
\lim_{\vec{k}_2\parallel\vec{p}_2}\Sfunm\,&=&1\,,
\label{Sfuncolm}\\
\lim_{k_2^0\to 0}\Sfunpm\,&\ne& 0\,.
\label{Sfunsoft}\\ \nonumber
\eeqn
The expectation value for any observable $O$ will then be written as
\beq
\VEV{O}=\VEV{O}_+ + \VEV{O}_-\,,
\eeq
where (see eqs.~(4.14)--(4.16) of ref.~\cite{Frixione:2002ik})
\beqn
\VEV{O}_\pm&=&\sum_{ab}\int dx_1 dx_2 d\phi_2\Bigg[
O(\Ktwo)\Sfunpm(\Ktwo)\frac{d\Sigma_{ab}^{(f)}}{d\phi_2}\evBB
+O(\Kone)\Sfunpm(\Kone)\frac{1}{\Ione}\left(\frac{d\Sigma_{ab}^{(b)}}{d\phi_1}
+\frac{d\Sigma_{ab}^{(sv)}}{d\phi_1}\right)
\nonumber \\*&& \phantom{d\phi_2\sum_{ab}\int dx_1 dx_2}
+O(\Kqone)\frac{1}{\Iqone}
\frac{d\Sigma_{ab}^{(c\pm)}}{d\phi_1 dx}\evBB
-O(\Kone)\frac{1}{\Iqone}
\frac{d\Sigma_{ab}^{(c\pm)}}{d\phi_1 dx}\cntBB
\nonumber \\*&& \phantom{d\phi_2\sum_{ab}\int dx_1 dx_2}
-\{O(\Kone)\Sfunpm(\Kone),O(\Kqone)\}\frac{d\Sigma_{ab}^{(f)}}{d\phi_2}\cntBB
\Bigg].
\label{nlosubtint}
\eeqn
As can be seen in ref.~\cite{Frixione:2002ik}, the $\Sigma$ terms in 
equation above are defined as the partonic short-distance cross sections, 
times the luminosity factors.

In writing eq.~(\ref{nlosubtint}), we have split (using the $\Sfun$ functions)
the Born ($\Sigma_{ab}^{(b)}$) and soft-virtual ($\Sigma_{ab}^{(sv)}$) 
contributions into two terms, and have associated them with 
the corresponding real-emission contributions (see 
refs.~\cite{Frixione:2005vw,Frederix:2009yq}). We have also used the 
collinear limits of eqs.~(\ref{Sfuncolp}) and~(\ref{Sfuncolm}), 
that imply that the $\Sfun$ functions
associated with the collinear counterterms are trivial. Finally, note
that the collinear remainders $\Sigma_{ab}^{(c\pm)}$ need not be multiplied
by the $\Sfun$ functions.

Equation~(\ref{nlosubtint}) must be further manipulated in order to use
it in MC@NLO; in particular, one has to apply the so-called event projection
which allows one to define a unique kinematic configuration associated
with all counterevents (for a given real-emission configuration)
-- see sect.~A.4 of ref.~\cite{Frixione:2002ik}. As discussed in that
paper, although event projection is largely arbitrary, in the context
of MC@NLO it turns out to be convenient to derive it from the behaviour
of the Monte Carlo one interfaces to. We shall therefore discuss in the
next subsection the behaviour of \PY\ relevant to this issue.

\subsection{Event projection with PYTHIA}
The way in which \PY\ deals with the simulation of $V$ production is the 
following~\cite{Sjostrand:2006za,Miu:1998ju}. First, the hard process is 
generated. 
This implies the generation of the two Bjorken $x$'s entering such a process, 
which we shall denote by $\xBpyo$ and $\xBpyt$ for the parton coming from
the left and from the right respectively. The partonic c.m. energy
squared is
\beq
s_0=\xBpyo\xBpyt S\,,
\label{spart}
\eeq
with $S$ the collider energy squared. In the case of single-$V$ production,
$s_0=\mV^2$, with $\mV$ the mass of $V$ (or its virtuality if lepton pair
production is considered), but eq.~(\ref{spart}) is obviously valid
regardless of the production process. \PY\ then begins the showers.
Choosing on statistical basis which leg emits ``first'' (such an emission 
is the only one that matters as far as MC@NLO is concerned), the shower
variables are related to the momenta given in eq.~(\ref{Rproc})
as follows:
\beqn
z&=&\frac{s_0}{(p_1 +p_2)^2}\,,
\label{shvz}
\\
t_i & = &(p_i -k_2)^2\,,
\label{shvQ2}
\eeqn
with $i=1,2$ for the emissions from the parton coming from the left and
from the right respectively. The branching generated with
eqs.~(\ref{shvz}) and~(\ref{shvQ2}) is such that the Bjorken $x$'s
associated with the initial-state partons of eq.~(\ref{Rproc})
(which we shall denote by $\hxBpyo$ and $\hxBpyt$) are
\beq
\hxBpyo=\xBpyo/z\,,\;\;\;\;\;\;\hxBpyt=\xBpyt
\eeq
for an emission from leg 1, and
\beq
\hxBpyo=\xBpyo\,,\;\;\;\;\;\;\hxBpyt=\xBpyt/z
\eeq
for an emission from leg 2. In the branching, the c.m. energy of
the Born-level subprocess is conserved (which, in the present case, is
equivalent to the requirement that the virtuality of $V$ be a constant),
and thus
\beq
s\,=\,(p_1+p_2)^2\,=\,\hxBpyo\hxBpyt S\,=\,s_0/z.
\label{ss0z}
\eeq
An easy way to achieve this is that of imposing that the Bjorken
$x$'s of the Born-level subprocesses, $\xBpyi$, be separately conserved.
\begin{figure}[htb]
  \begin{center}
    \epsfig{figure=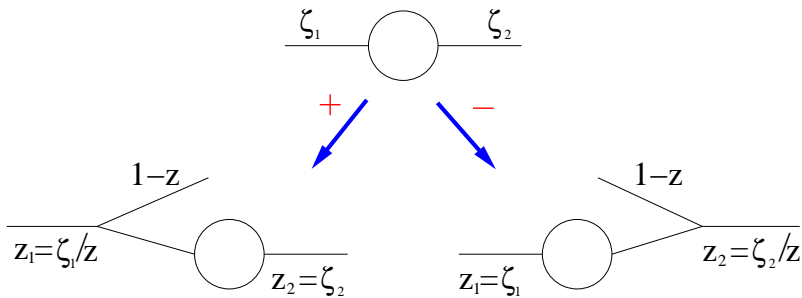,width=0.7\textwidth}
\caption{\label{fig:MCbran} 
Assignments of momentum fractions in the case of initial-state branchings
as done in \PY. Branchings from legs 1 and 2 are denoted in MC@NLO by
$+$ and $-$ respectively. The blob represents the final state at the
Born level.
}
  \end{center}
\end{figure}

As shown in ref.~\cite{Frixione:2002ik}, the event-projection procedure
can be formally determined by constructing two ``observables'' that
are conserved in the branching process. According to what has been
discussed above, a possible choice is:
\beqn
O_1&=&\xBpyo\xBpyt S\,,
\label{Obs1}
\\*
O_2&=&\frac{1}{2}\log\frac{\xBpyo}{\xBpyt}\,.
\label{Obs2}
\eeqn
It is worth noting that while eq.~(\ref{Obs1}) is the same as
in \HW, eq.~(\ref{Obs2}) does not coincide with either of the \HW\
choices considered in ref.~\cite{Frixione:2002ik}; the implication
of this fact is that we obtain two different event projections,
depending on whether it is leg 1 or 2 that emits. We stress that
eqs.~(\ref{Obs1}) and~(\ref{Obs2}), and the resulting manipulations
we are now going to describe, do not depend on the nature of the 
Born-level final state (a $V$ boson rather than -- say -- a three-jet
configuration), owing to the fact that \PY\ adopts the so-called
$s$-approach~\cite{Sjostrand:2006za} when doing initial-state 
branchings for all hadroproduction processes. In the context of the
$s$-approach, the invariant mass ($\sqrt{s_0}$) of the final-state system
at the Born level is kept constant during the branching. This allows
one to formally replace $V$ with the set of final-state particles at 
the Born level, and $k_1$ with the sum of their four momenta, which does not
entail any changes to eqs.~(\ref{spart})--(\ref{ss0z}), from which we 
ultimately derive eqs.~(\ref{Obs1}) and~(\ref{Obs2}). The procedure
\begin{figure}[htb]
  \begin{center}
    \epsfig{figure=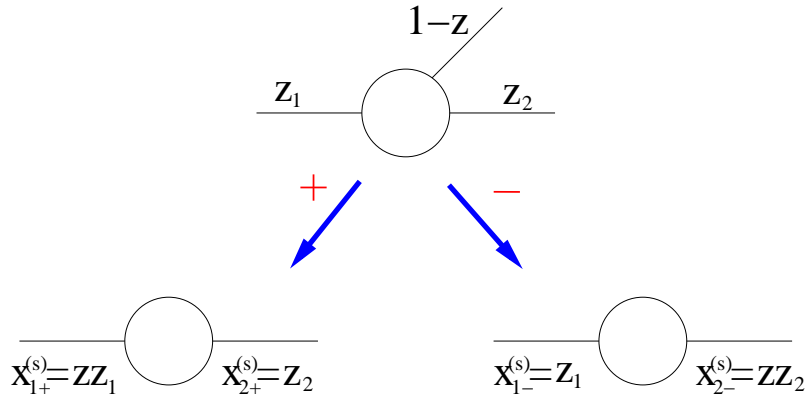,width=0.7\textwidth}
\caption{\label{fig:EvPr} 
Assignments of momentum fractions according to the event-projection
procedure, for the soft configurations. See the text for details.
}
  \end{center}
\end{figure}
described here is thus fully general, and is not restricted to $V$-boson
production. We recall that event projection is a way to re-write
the soft and collinear counterterms in a pure-NLO computation, so as
all counterterms associated with a given $\Sfun$ contribution at the
real-emission level (in the present case, $\Sfunp d\Sigma^{(f)}$ or
$\Sfunm d\Sigma^{(f)}$) have the same kinematics. In order to achieve
this, it turns out to be convenient to define the counterterm kinematics
starting from a fixed real-emission kinematics. This situation
is depicted in fig.~\ref{fig:EvPr} for soft counterterms (the case of
collinear counterterms is essentially identical, the only differences being 
in the assignments of the momentum fractions of the incoming partons, which 
we shall specify in what follows). Since the event projections we define here 
are ultimately motivated by the branching strategy of \PY, the procedure of 
fig.~\ref{fig:EvPr} is by construction the inverse of that depicted in 
fig.~\ref{fig:MCbran}. We shall return to this point in sect.~\ref{sec:xsec}.

In order to actually obtain event projections, we use the 
master equations (A.36) and (A.37) of ref.~\cite{Frixione:2002ik}. 
For an emission from leg 1, these equations read in our case
\beqn
O_1(\Ktwo(\hxBpyo,\hxBpyt,\phi_2))&=&
O_1(\Kone_s(\xosp,\xtsp,\phi_2^{(s)}))\,,
\label{Oonecond}
\\
O_2(\Ktwo(\hxBpyo,\hxBpyt,\phi_2))&=&
O_2(\Kone_s(\xosp,\xtsp,\phi_2^{(s)}))\,,
\label{Otwocond}
\eeqn
and their analogues for the collinear case (where one formally
replaces $s$ with $c+$ in the r.h.s. of eqs.~(\ref{Oonecond})
and~(\ref{Otwocond})). Note that, having treated separately the
two collinear singularities in the NLO cross section, the case
$c-$ need not be considered when studying branching from leg 1). 
We find:
\beqn
&&
\xosp=z\,\hxBpyo\,,\phantom{aaaaaaaaa}
\xtsp=\hxBpyt\,,
\label{xtilsp}
\\&&
\xocpp=\xosp/z\,,\phantom{aaaaaa}
\xtcpp=\xtsp\,.
\label{xtilcpp}
\eeqn
For an emission from leg 2, we find instead
\beqn
&&
\xosm=\hxBpyo\,,\phantom{aaaaaaaaaa\,}
\xtsm=z\,\hxBpyt\,,
\label{xtilsm}
\\&&
\xocmm=\xosm\,,\phantom{aaaaaaaa}
\xtcmm=\xtsm/z\,.
\label{xtilcmm}
\eeqn
At this point, following what was done in sect.~4.4 of
ref.~\cite{Frixione:2002ik}, one uses eqs.~(\ref{xtilsp}) and~(\ref{xtilcpp})
to perform the event projection transformation of $\VEV{O}_+$, and 
eqs.~(\ref{xtilsm}) and~(\ref{xtilcmm}) to deal with $\VEV{O}_-$.
Once event projection has been achieved, the MC@NLO short distance
cross section can be defined by including the MC subtraction terms.
We shall do this in the next subsection.

\subsection{MC subtraction terms}
At the NLO in $\as$, the cross section resulting from \PY\ is
\beq
d\sigma\xMCB=d\sigma^{(+)}\xMCB+d\sigma^{(-)}\xMCB\,,
\eeq
where the two terms on the r.h.s. account for emissions from leg 1
and 2 respectively. They read:
\beqn
d\sigma^{(+)}\xMCB&=&\sum_{abc}d\xBpyo d\xBpyt \frac{\as}{2\pi}
d\sigma_{ab}^{(b)}(\xBpyo P_1,\xBpyt P_2)
\frac{dt_1}{t_1}\frac{dz}{z}
P^{(0)}_{ac}(z)f_c^{\Hone}(\xBpyo/z)f_b^{\Htwo}(\xBpyt),\phantom{aaa}
\label{sigmaMCp}
\\
d\sigma^{(-)}\xMCB&=&\sum_{abc}d\xBpyo d\xBpyt \frac{\as}{2\pi}
d\sigma_{ab}^{(b)}(\xBpyo P_1,\xBpyt P_2)
\frac{dt_2}{t_2}\frac{dz}{z}
P^{(0)}_{bc}(z)f_a^{\Hone}(\xBpyo)f_c^{\Htwo}(\xBpyt/z).\phantom{aaa}
\label{sigmaMCm}
\eeqn
Following what was done in sect.~A.5 of ref.~\cite{Frixione:2002ik}, we can
manipulate the equations above to render them suitable for an integration
together with the NLO parton-level cross section. First of all, one notes
that the following identifications are valid owing to the construction
of event projections:
\beq
\xBpyo\equiv\xosp\,,\;\;\;\;\;
\xBpyt\equiv\xtsp\,,
\eeq
in eq.~(\ref{sigmaMCp}), and
\beq
\xBpyo\equiv\xosm\,,\;\;\;\;\;
\xBpyt\equiv\xtsm\,,
\eeq
in eq.~(\ref{sigmaMCm}). We can therefore perform a change of integration
variables:
\beq
d\sigma^{(\pm)}\xMCB=\sum_{ab}dz_1 dz_2 d\bSigma_{ab}^{(\pm)}\xMCB\,,
\label{sigmaMCt}
\eeq
with
\beqn
d\bSigma_{ab}^{(+)}\xMCB&=&
\frac{1}{z}\frac{\partial(\xosp,\xtsp)}{\partial(z_1,z_2)}
f_a^{\Hone}(\xosp/z)f_b^{\Htwo}(\xtsp)d\sigma_{ab}^{(+)}\xMCB
\\*
d\bSigma_{ab}^{(-)}\xMCB&=&
\frac{1}{z}\frac{\partial(\xosm,\xtsm)}{\partial(z_1,z_2)}
f_a^{\Hone}(\xosm)f_b^{\Htwo}(\xtsm/z)d\sigma_{ab}^{(-)}\xMCB
\label{bSigmaMCdef}
\eeqn
and
\beqn
\frac{d\sigma_{ab}^{(+)}}{d\phi_2}\xMCBB&=&
\sum_c \frac{8\pi\as}{\mV^2 \hat{t}_1}  
P_{ca}^{(0)}(z)\mborn_{cb}(\xosp P_1,\xtsp P_2)\stepf_+\,,
\label{dsigMCp}
\\*
\frac{d\sigma_{ab}^{(-)}}{d\phi_2}\xMCBB&=&
\sum_c \frac{8\pi\as}{\mV^2 \hat{t}_2}  
P_{cb}^{(0)}(z)\mborn_{ac}(\xosm P_1,\xtsm P_2)\stepf_-\,.
\label{dsigMCm}
\eeqn
Here, we have introduced the rescaled virtualities
\beq
\hat{t}_i=-t_i/\mV^2\,.
\eeq
Furthermore, we have explicitly included the factors $\stepf_\pm$,
which are related to the choice of the maximum virtuality allowed
during shower evolution. We have e.g.
\beq
\abs{t_i}\le\mV^2\;\;\;\;\Longrightarrow\;\;\;\;
\stepf_\pm=\stepf\!\left(y-\frac{1-3x}{1-x}\right),
\label{maxvirt}
\eeq
with $y$ being the relevant angular variable associated with the
FKS parton in the FKS subtraction (here the cosine of the angle between
$\vec{k}_2$ and $\vec{p}_i$ for emissions from leg $i$), and $1-x$ the
(normalized) energy of the FKS parton. In the case of \PY, at variance
with \HW, $x\equiv z$.

\subsection{MC@NLO short-distance cross sections\label{sec:xsec}}
At this point, we have all the ingredients needed to generate hard-subprocess
events according to the MC@NLO formalism, which will be subsequently showered
by \PY. Recalling that we denote by $\bSigma^{(\alpha)}$ the contributions
$\Sigma^{(\alpha)}$ to the NLO parton-level cross sections after event
projection, event generation will be performed following the procedure
outlined in sect.~4.5 of ref.~\cite{Frixione:2002ik} (see also
sect.~3 of ref.~\cite{Frixione:2005vw}), using the following integrals:
\beqn
I_{\clH}^{(\pm)}&=&\sum_{ab}\int dz_1 dz_2 d\phi_2
\Bigg(\Sfunpm\frac{d\bSigma_{ab}^{(f)}}{d\phi_2}\evBB
-\frac{d\bSigma_{ab}^{(\pm)}}{d\phi_2}\xMCBB\Bigg)\,,
\label{Ithreedef}
\\
I_{\clS}^{(\pm)}&=&\sum_{ab}\int dz_1 dz_2 d\phi_2 
\Bigg[-\Sfunpm\frac{d\bSigma_{ab}^{(f)}}{d\phi_2}\cntBB
+\frac{d\bSigma_{ab}^{(\pm)}}{d\phi_2}\xMCBB
\nonumber \\*&& \phantom{a}
+\frac{\Sfunpm}{\Ione}\Bigg(\frac{d\bSigma_{ab}^{(b)}}{d\phi_1}
+\frac{d\bSigma_{ab}^{(sv)}}{d\phi_1}\Bigg)
+\frac{1}{\Iqone}\Bigg(
\frac{d\bSigma_{ab}^{(c\pm)}}{d\phi_1 dx}\evBB
-\frac{d\bSigma_{ab}^{(c\pm)}}{d\phi_1 dx}\cntBB\Bigg)
\Bigg]\,.
\label{Itwodef}
\eeqn
We remind the reader that all integrals defined in these equations
are separately finite, and that
\beq
\sigma_{tot}=I^{(+)}_{\clS}+I^{(+)}_{\clH}+I^{(-)}_{\clS}+I^{(-)}_{\clH}
\eeq
is an exact equation (with $\sigma_{tot}$ the fully-inclusive
NLO rate).

Although eqs.~(\ref{Ithreedef}) and~(\ref{Itwodef}) solve the problem
of the generation of hard-subprocess events, a further simplification is 
possible. We start by considering the integrals $I_{\clH}^{(\pm)}$, 
that are used to 
generate $\clH$ events. These two integrals have to be computed separately.
However, one observes that, for a given choice of \mbox{$(z_1,z_2,\phi_2)$},
the kinematics associated with these two contributions are actually
identical, and the integrals can therefore be computed together.
This is equivalent to using the following integral for the generation
of $\clH$ events:
\beq
I_{\clH}\equiv I_{\clH}^{(+)}+I_{\clH}^{(-)}=
\sum_{ab}\int dz_1 dz_2 d\phi_2
\Bigg(\frac{d\bSigma_{ab}^{(f)}}{d\phi_2}\evBB
-\sum_{L=\pm}\frac{d\bSigma_{ab}^{(L)}}{d\phi_2}\xMCBB\Bigg)\,.
\label{Ithreedef2}
\eeq
The first term in the integrand has been simplified thanks to
eq.~(\ref{Sfunnorm}). For the computation of the second term in
the integrand we stress that, as eq.~(\ref{Ithreedef2}) explicitly
indicates, $z_i$ are the integration variables. Therefore, when we
evaluate eqs.~(\ref{dsigMCp}) and~(\ref{dsigMCm}), the variables
$x_{i\pm}^{(s)}$ have to be computed according to eqs.~(\ref{xtilsp})
and~(\ref{xtilsm}). From the physical viewpoint, this implies that
the two Born matrix elements appearing implicitly in $\bSigma_{ab}^{(\pm)}$ 
are evaluated in two different kinematics configurations, that eventually 
result after the first branching in the same real-emission kinematics
\mbox{$(z_1,z_2,\phi_2)$}. This situation is precisely the one depicted
in fig.~\ref{fig:EvPr}.

A simplification analogous to that of eq.~(\ref{Ithreedef2}) is 
also possible in the case of $\clS$ events,
although the argument is slightly more involved. If one fixes
\mbox{$(z_1,z_2,\phi_2)$} (i.e., the real-emission kinematics),
the kinematic configurations associated with $I_{\clS}^{(+)}$ and
$I_{\clS}^{(-)}$ in eq.~(\ref{Itwodef}) are different, owing to the
fact that the two integrals are computed using different event
projections. Formally, one can introduce two mappings, $\EVprjmapp$ 
and $\EVprjmapm$, representing these event projections; for a
given real-emission kinematic configuration $\Ktwo$, the two
Born-like configurations associated with $I_{\clS}^{(+)}$ and
$I_{\clS}^{(-)}$ can be denoted by
\beq
\Kone^+=\EVprjmapp\,\Ktwo\,,\;\;\;\;\;\;
\Kone^-=\EVprjmapm\,\Ktwo
\label{KonefKtwo}
\eeq
respectively; $\Kone^+$ and $\Kone^-$ are the hard configurations to be
showered by \PY. The $\EVprjmappm$ mappings are pictorially represented in
fig.~\ref{fig:EvPr} by the two thick arrows, with $\Ktwo$, $\Kone^+$,
and $\Kone^-$ the configurations depicted in the upper part, lower left
corner, and lower right corner of that figure respectively.
On the other hand, one can fix a Born-like configuration
$\Kone$ and, using the inverse of the maps $\EVprjmappm$ (which we
denote by $\IEVprjmappm$), work out the
real-emission configurations to be used in the computation of the
integrand of $I_{\clS}^{(\pm)}$:
\beq
\Ktwo^+=\IEVprjmapp\,\Kone\,,\;\;\;\;\;\;
\Ktwo^-=\IEVprjmapm\,\Kone\,.
\label{invmap}
\eeq
As we have already stressed, although there is an ample freedom
in choosing $\EVprjmappm$, it is best to adopt a form motivated
by what the PSMC does when branching. In practice, it is therefore convenient
first to obtain $\IEVprjmappm$ from the PSMC, and then (by inverting them) 
$\EVprjmappm$. In the case of \PY, $\IEVprjmappm$ amount to performing
a boost of the Born-level four momenta in the transverse direction,
to balance the transverse momentum of the parton produced in the
branching ($c(k_2)$ in eq.~(\ref{Rproc})), followed by a boost in
the longitudinal direction, to e.g. the lab frame. The information
on the latter is equivalent to the assignments of the momentum fractions
of the incoming particles. The prescription for the transverse boost
given above is obviously trivial in the case of $V$ production
(since $k_{1{\sss T}}=-k_{2{\sss T}}$), but can be applied to final
states with arbitrary multiplicity. Thus, what done here is also
valid for final states more complicated than single $V$.

The two procedures related to eq.~(\ref{KonefKtwo}) and~(\ref{invmap})
are equivalent, since the integrals are performed over 
the whole phase space (possibly subject to kinematic restrictions,
which are however identical in the two cases). Equation~(\ref{invmap}) 
corresponds to the situation depicted in fig.~\ref{fig:MCbran},
except for the seemingly different assignments of the momentum
fractions of the incoming particles. These differences are however 
immaterial, since such fractions are simply integration variables
which can be manipulated and renamed at will. We shall now proceed
to perform such manipulations, starting from a change of variables 
in eq.~(\ref{Itwodef})
\beq
(z_1,z_2,\phi_2)\;\;\longrightarrow\;\;(\xospm,\xtspm,\phi_2^\pm)\,.
\label{vchange}
\eeq
Furthermore, consistently with eq.~(\ref{invmap}), the integrands have to be 
computed with the suitable kinematic configurations, $\Ktwo^+$ or $\Ktwo^-$. 
We can take these two operations into account by the formal replacements
\beq
\bSigma_{ab}^{(\alpha)}(\Ktwo(z_i))\;\;\longrightarrow
\hSigma_{ab}^{(\alpha)}(\Ktwo^\pm(x_{j\pm}^{(s)}))=
\frac{\partial(z_1,z_2)}{\partial(\xospm,\xtspm)}
\bSigma_{ab}^{(\alpha)}(\Ktwo(z_i(x_{j\pm}^{(s)})))\,.
\label{bStohS}
\eeq
It is easy to realize that, in the case of the Born, soft-virtual,
and soft counterterms contributions, we have by construction
\beq
\hSigma_{ab}^{(\alpha)}=\Sigma_{ab}^{(\alpha)}\,,
\label{bStohSsoft}
\eeq
since for these terms the jacobian appearing in eq.~(\ref{bStohS}) exactly 
compensates the one in the definition of the $\bSigma_{ab}^{(\alpha)}$
terms, see eq.~(4.18) of~\cite{Frixione:2002ik}. 
In general, this is not true for contributions with a purely collinear
structure (such as the collinear counterterms to the real-emission matrix
elements). Equations~(\ref{vchange}), (\ref{bStohS}), and~(\ref{bStohSsoft}) 
allow us to rewrite
\beqn
I_{\clS}^{(\pm)}&=&\sum_{ab}\int d\xospm d\xtspm d\phi_2^\pm
\Bigg[-\Sfunpm\frac{d\hSigma_{ab}^{(f)}}{d\phi_2}\cntBB
+\frac{d\hSigma_{ab}^{(\pm)}}{d\phi_2}\xMCBB
\nonumber \\*&& \phantom{a}
+\frac{\Sfunpm}{\Ione}\Bigg(\frac{d\Sigma_{ab}^{(b)}}{d\phi_1}
+\frac{d\Sigma_{ab}^{(sv)}}{d\phi_1}\Bigg)
+\frac{1}{\Iqone}\Bigg(
\frac{d\hSigma_{ab}^{(c\pm)}}{d\phi_1 dx}\evBB
-\frac{d\hSigma_{ab}^{(c\pm)}}{d\phi_1 dx}\cntBB\Bigg)
\Bigg]\,.
\label{Itwodef2}
\eeqn
At this point, the two integrals in eq.~(\ref{Itwodef2}) correspond by
construction to the same Born-like kinematic configuration (to be showered
by \PY), and can therefore be integrated together. It is thus convenient
to {\em rename} the integration variables
\beq
(\xospm,\xtspm,\phi_2^\pm)\;\;\longrightarrow\;\;
(\xBpyo,\xBpyt,\phi_2)\,,
\eeq
and use
\beq
I_{\clS}\equiv I_{\clS}^{(+)}+I_{\clS}^{(-)}
\label{ISsum}
\eeq
for the generation of $\clS$ events. One observes that in the integrand
of eq.~(\ref{ISsum}) there will be a term
\beq
\sum_{L=\pm}\frac{\SfunL}{\Ione}\Bigg(\frac{d\Sigma_{ab}^{(b)}}{d\phi_1}
+\frac{d\Sigma_{ab}^{(sv)}}{d\phi_1}\Bigg)=
\frac{1}{\Ione}\Bigg(\frac{d\Sigma_{ab}^{(b)}}{d\phi_1}
+\frac{d\Sigma_{ab}^{(sv)}}{d\phi_1}\Bigg)\,,
\eeq
where we have made use of eq.~(\ref{Sfunnorm}). The same kind of
simplification will occur in the sum of the two soft counterterms.
As far as the collinear counterterms are concerned, we stress that
the $\Sfunpm$ terms that multiply them are equal to one (by construction
of the $\Sfun$ functions -- see eqs.~(\ref{Sfuncolp}) and~(\ref{Sfuncolm})).
We arrive therefore at the following form:
\beqn
I_{\clS}&=&\sum_{ab}\int d\xBpyo d\xBpyt d\phi_2 
\Bigg[-\frac{d\hSigma_{ab}^{(f)}}{d\phi_2}\cntBB
+\sum_{L=\pm}\frac{d\hSigma_{ab}^{(L)}}{d\phi_2}\xMCBB
\nonumber \\*&& \phantom{a}
+\frac{1}{\Ione}\Bigg(\frac{d\Sigma_{ab}^{(b)}}{d\phi_1}
+\frac{d\Sigma_{ab}^{(sv)}}{d\phi_1}\Bigg)
+\frac{1}{\Iqone}\sum_{L=\pm}\Bigg(
\frac{d\hSigma_{ab}^{(cL)}}{d\phi_1 dx}\evBB
-\frac{d\hSigma_{ab}^{(cL)}}{d\phi_1 dx}\cntBB\Bigg)
\Bigg]\,.
\label{Itwodef3}
\eeqn
Equations~(\ref{Ithreedef2}) and~(\ref{Itwodef3}) are our final expressions,
used for the generation of $\clH$ and $\clS$ events respectively. The net
result of the various manipulations carried out in this section is that
it is still possible to treat simultaneously the two initial-state
collinear singularities, as was the case when matching with \HW.
We point out that the fact that eqs.~(\ref{Ithreedef2}) and~(\ref{Itwodef3}) 
do not depend upon the $\Sfun$ functions is not a property of $V$-boson
production, but applies as well to all processes whose Born-level final-state
particles are all colour singlets\footnote{It is also possible, but not
mandatory, to treat in this way processes in which {\em all} 
strongly-interacting final-state particles are massive, as e.g. in 
$t\bar{t}$ production. See ref.~\cite{Frixione:2003ei} for an explicit 
example.}. In fact, for this independence of $\Sfun$ functions to occur, 
we only need eq.~(\ref{Sfunnorm}) to hold, which is
true in the cases just mentioned (see ref.~\cite{Frederix:2009yq} for
further details on the construction of $\Sfun$ functions). Clearly,
in order to make use of eq.~(\ref{Sfunnorm}), $\Sfunp$ and $\Sfunm$ 
need be computed for the same kinematic configuration. This is in fact
what happens, in spite of the fact that in the intermediate steps of
the event-projection procedure we had to treat separately the $+$ and 
$-$ contributions. For $\clH$ events this is trivially true by 
construction -- both $I_{\clH}^{(\pm)}$ in eq.~(\ref{Ithreedef})
are associated with the kinematic configuration denoted by $\Ktwo$ in
eq.~(\ref{KonefKtwo}) (i.e. the upper part of fig.~\ref{fig:EvPr}). 
On the other hand, the short-distance cross sections for $\clS$ 
events we started from, eq.~(\ref{Itwodef}), are associated with  
two different kinematic configurations, denoted by $\Kone^\pm$ 
in eq.~(\ref{KonefKtwo}) (i.e. the lower part of fig.~\ref{fig:EvPr}). 
Thanks to the procedure described above, we have manipulated 
eq.~(\ref{Itwodef}) precisely to be able to associate the two $\pm$ 
contributions with the same kinematic configuration, denoted by $\Kone$ 
in eq.~(\ref{invmap}) (i.e. the upper part of fig.~\ref{fig:MCbran}). 
Owing to the properties of event projections that we have discussed above,
the whole procedure can obviously be carried out for any final-state
multiplicity, with $\Ktwo$ and $\Kone$ formally replaced by the
real-emission and Born-level configurations respectively.
Finally, we stress that the manipulations performed here will also
be valid in the case of strongly-interacting particles in the final
state. The only difference is that, in such a case, eq.~(\ref{Sfunnorm}) 
will not hold any longer, and therefore the quantity \mbox{$\Sfunp+\Sfunm$}
will appear in eqs.~(\ref{Ithreedef2}) and~(\ref{Itwodef3}) as a factor
multiplying the real-emission, soft counterterms, soft-virtual,
and Born contributions (on the other hand, the purely collinear terms
will be unchanged, owing to the fact that eqs.~(\ref{Sfuncolp})
and~(\ref{Sfuncolm}) hold regardless of the nature of the final state).
Clearly, the contributions due to branchings of Born-level final-state
strongly-interacting particles will be given by short-distance cross
sections analogous to those of eqs.~(\ref{Ithreedef2}) and~(\ref{Itwodef3})
(see e.g. sect.~3 of ref.~\cite{Frixione:2005vw}).
The computation of such contributions is beyond the scope of the
present work, and we postpone it to a future publication.

We conclude this section by stressing that, due to the form of 
the shower variables used by \PY, the soft limit of the
sum of the MC subtraction terms coincides with that of the real-emission
matrix elements, which was not the case for \HW. This implies that
here we can set ${\cal G}\equiv 1$, where ${\cal G}$ is the function
introduced in sect.~A.5 of ref.~\cite{Frixione:2002ik} -- we refer
the reader to that paper for a discussion on this issue.

\section{Results\label{sec:res}}
In this section, we present sample results relevant to $W^+$ production
in $pp$ collisions at \mbox{$\sqrt{S}=14$~TeV}. Our aim is not that
of performing a phenomenological study, but rather that of presenting 
a few control plots that show that the matching of the NLO results
for single vector boson production with \PY\ according to the
MC@NLO formalism works as we expect. This is non trivial, given
the differences between \HW\ and \PY, which in turn result in
different short-distance MC@NLO cross sections, as discussed previously.
We set $m_W=80.4$~GeV, $\Gamma_W=2.14$~GeV, and use
CTEQ6.6~\cite{Nadolsky:2008zw} PDFs. Our default scale choices
are $\muF=\muR=\mt$, where $\mt$ is the transverse mass of the $W$.
When reconstructing jets, we adopt
the $k_T$-jet-finding algorithm of ref.~\cite{Catani:1993hr}, with 
\mbox{$Y_{\sss\rm cut}=(10~{\rm GeV})^2$}.

\begin{figure}[htb]
  \begin{center}
    \epsfig{figure=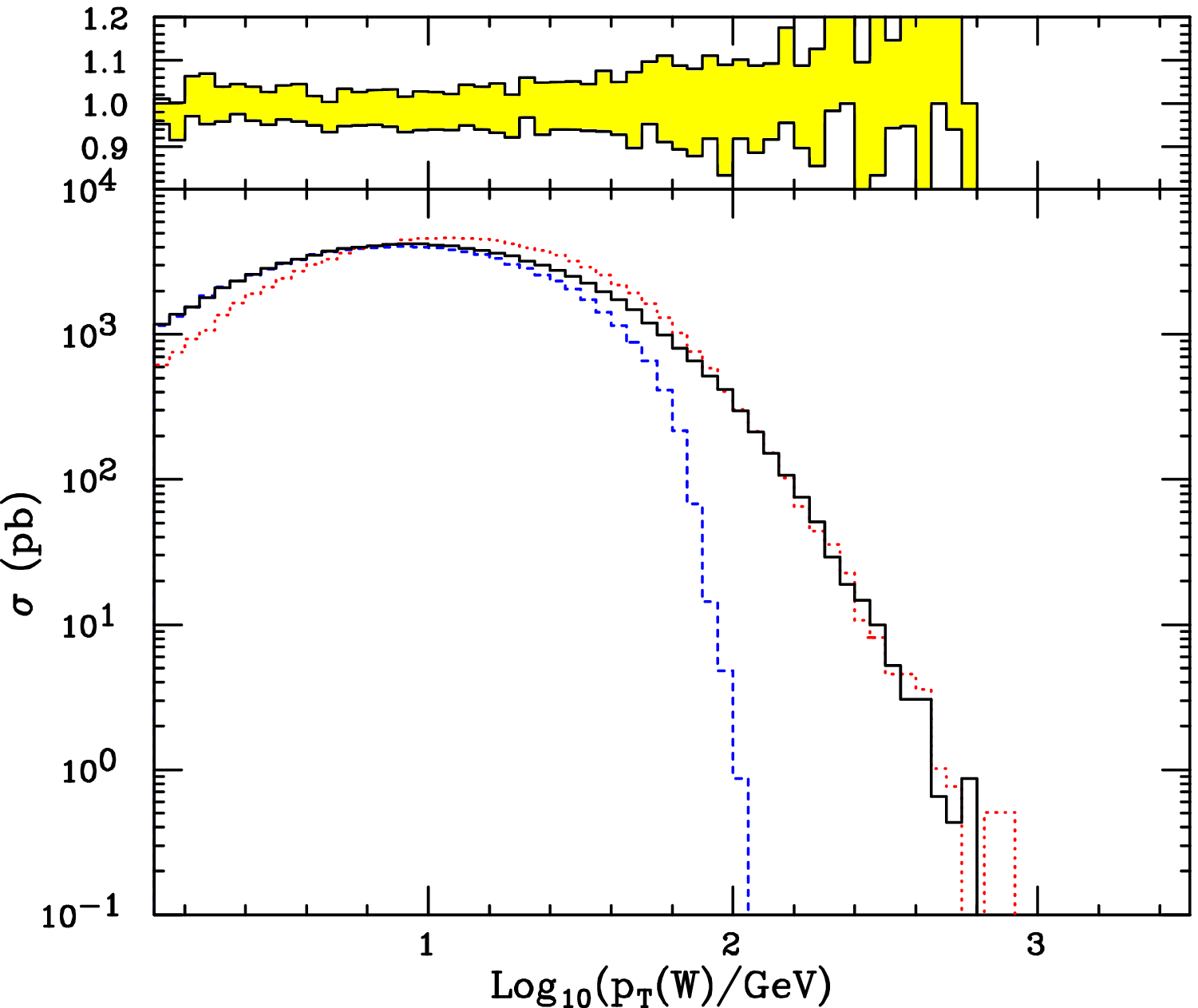,width=0.49\textwidth}
    \epsfig{figure=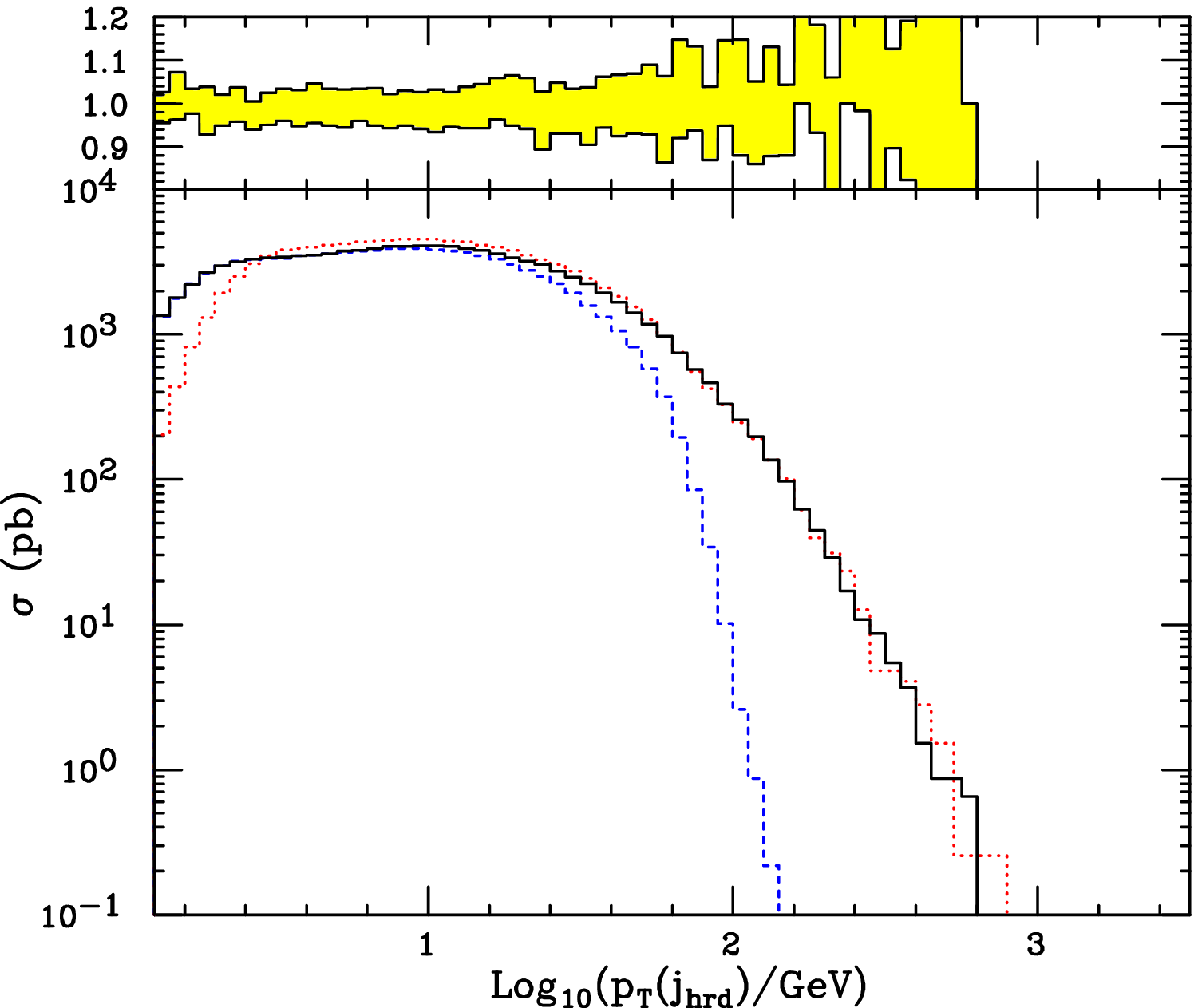,width=0.49\textwidth}
\caption{\label{fig:pt} 
Solid (black): \NLPY. Dotted (red): \NLHW. Dashed (blue):
\PY\ standalone (rescaled), without matrix element corrections.
Left pane: $\pt$ of the $W$ boson. Right pane: $\pt$ of the
hardest jet of the event. The insets show the fractional
scale dependence of \NLPY, computed as described in the text.
}
  \end{center}
\end{figure}

\begin{figure}[htb]
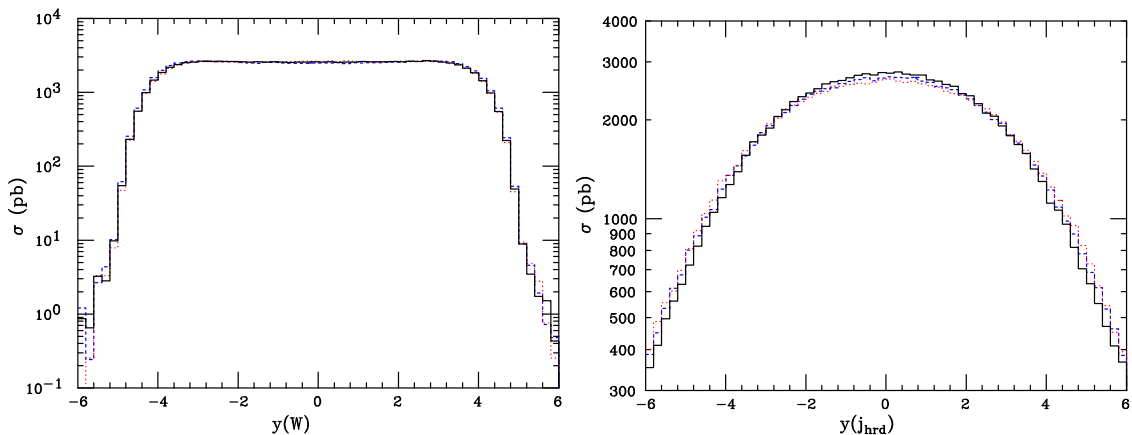

  \begin{center}
    \epsfig{figure=pysvb_yV.eps,width=0.49\textwidth}
    \epsfig{figure=pysvb_yjhrd.eps,width=0.49\textwidth}
\caption{\label{fig:y} 
Same as in fig.~\ref{fig:pt}. Left pane: rapidity of the $W$ boson.
Right pane: rapidity of the hardest jet of the event.
}
  \end{center}
\end{figure}

\begin{figure}[htb]
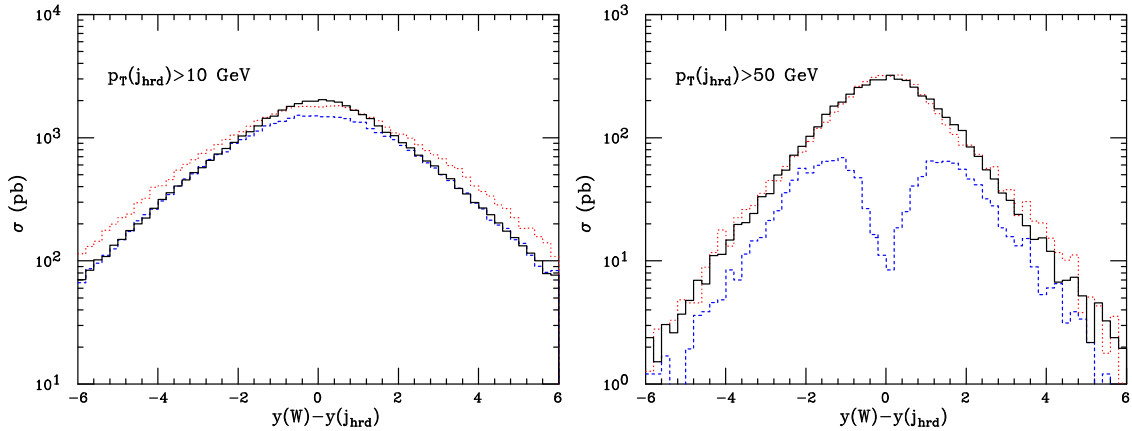

  \begin{center}
    \epsfig{figure=pysvb_Dely_pt10.eps,width=0.49\textwidth}
    \epsfig{figure=pysvb_Dely_pt50.eps,width=0.49\textwidth}
\caption{\label{fig:Dely} 
Same as in fig.~\ref{fig:pt}, for the difference in rapidity between
the $W$ boson and the hardest jet of the event. Two cuts on the
$\pt$ of the jet are considered.
}
  \end{center}
\end{figure}

In each of the plots we present in figs.~\ref{fig:pt}--\ref{fig:Dely}, 
we display three histograms.
The solid (black) histograms are the results of MC@NLO with \PY\
(which we shall call \NLPY) i.e. what has been computed in this paper. 
The dotted (red) histograms are the results of MC@NLO with \HW\
(which we shall call \NLHW). Finally, the dashed (blue) histograms
are the results obtained with \PY\ standalone. 
When using \PY\ for showering hard events, we set MSTP(81)=0 and 
MSTP(91)=0, which corresponds to switching off multiple interactions 
and primordial $\kt$ respectively. We switch off matrix element corrections
by setting MSTP(68)=0, which also forces the maximum virtuality in the
shower to be equal to the vector boson mass, when standalone generation
is performed. On the other hand, hard subprocess events generated by \NLPY\ 
are given to \PY\ in the standard Les Houches format~\cite{Boos:2001cv}; 
we set PARP(67)=1 and {\tt SCALUP}=$m_W$ to have the same maximum scale in
the shower as in the standalone generation.
In order to facilitate the visual 
comparisons between \NLPY\ and \PY, the results of the latter are 
rescaled (by different factors, depending on the observables considered).
As has already been discussed at length in the literature, 
we expect MC@NLO to be identical in shape to the
PSMC results in the regions of the phase space dominated by those
large logarithms the PSMC is able to resum; this is the motivation for
comparing \NLPY\ with \PY\ standalone. Also, we expect MC@NLO to coincide,
in shape and normalization, with the NLO results in the regions where
hard emissions are dominant. Given that it has already been shown in the
past that this is the case for \NLHW, here we compare \NLPY\  directly with
\NLHW, rather than with the pure NLO results, since this will also
give us the opportunity to observe the different behaviours of the
two underlying PSMCs in the soft/collinear regions.
As far as negative weights are concerned, for $W$ production at the LHC 
their fraction with \NLPY\ is about 0.6\%, while in the case of \NLHW\ is
about 8\%. Since the phase-space parametrizations used in the two codes
are identical, this large difference is essentially due to the different 
choices of shower variables made by the two PSMCs.

In fig.~\ref{fig:pt} we consider the transverse momentum distributions
of the $W^+$ boson (left pane), and of the hardest jet of the event
(right pane). These distributions are interesting since in the large-$\pt$
region they are dominated by hard emissions, whereas in the small-$\pt$
one there are logarithmically-enhanced terms that need be resummed
(the pure NLO results diverge there). The comparisons among the three
results for the two distributions follow the same pattern, as we expect.
At small $\pt$'s, the shapes of \NLPY\ and of \PY\ are identical.
To show this clearly, the latter results have been rescaled so as
their first bins coincide with those resulting from \NLPY. On the
other hand, the \NLPY\ and \NLHW\ results are quite different in
this $\pt$ region, owing to the different treatment of soft and collinear
emissions by the two PSMCs. It is known that \PY\ tends to give softer
spectra than \HW, which is therefore what we expect to and do find 
with \NLPY\ and \NLHW. At large  $\pt$'s, \NLPY\ coincides with \NLHW\ 
-- both in shape and in absolute normalization, which is again what we expect.

In fig.~\ref{fig:pt} we also show the scale dependence of \NLPY, determined 
according to the following procedure. The renormalization-scale variations 
are defined as the differences between the results obtained with 
$(\muR,\muF)=(\mt,\mt)$ (i.e. the default),
and those obtained with $(\muR,\muF)=(f\mt,\mt)$, where $f=1/2$ and $f=2$;
these differences are computed bin-by-bin for all observables studied.
Likewise, for the factorization-scale variations one considers
$(\muR,\muF)=(\mt,f\mt)$. The renormalization- and factorization-scale
variations of like sign are then summed in quadrature, and the results
are then summed to (for positive variations) or subtracted from (for 
negative variations) the default cross section. The ratios of
the two predictions obtained in this way over the default cross section
are displayed in the insets of fig.~\ref{fig:pt} as the upper and
lower bounds of the shaded areas, for the two
observables considered there. In spite of the lack of statistics in
the high-$\pt$ tails\footnote{MC@NLO outputs unweighted events. The plots
presented here have been obtained with $5\mydot 10^5$-event samples, and
therefore less than five hundred events have $\pt$'s larger than a 
few hundreds GeV.} the trend is clear: the fractional scale dependence
grows from about $\pm 5$\% at low $\pt$'s to about $\pm 10$\% at large
$\pt$'s. This has to be compared with the pure-NLO result (not shown
here) that features a decrease in the scale dependence, from $\pm 15$\%
at low $\pt$'s to $\pm 10$\% at large $\pt$'s; this behaviour results
from a dependence on $\muF$ almost constant w.r.t. $\pt$, and a dependence 
on $\muR$ decreasing with $\pt$ (due to the running of $\as$).
The scale dependences of \NLPY\ and of the pure-NLO
predictions are therefore consistent at large $\pt$'s. This is what we
expect, since there the \NLPY\ result is basically coincident with
the NLO one, the effect of the shower being negligible on these inclusive
variables. The size of the scale dependence is also compatible with the
fact that, in fixed-order perturbation theory, ${\cal O}(\as)$ is actually
the first order that contributes to $\pt>0$. This observation therefore 
applies also to low $\pt$'s in the case of the pure-NLO predictions,
but it does not in the case of \NLPY. In fact, \NLPY\ fills the low- and
intermediate-$\pt$ regions mostly through the showering of $\clS$ events.
The shower, however, determines only the kinematics of the final-state
configuration (i.e., the $\pt$ of the $W$ and of the hardest jet
here), but the weights are given by the short-distance cross section
of eq.~(\ref{Itwodef3}), and therefore receive both ${\cal O}(\as^0)$
and ${\cal O}(\as)$ contributions. This implies that in the
low-$\pt$ region one expects \NLPY\ to have a scale dependence 
smaller than that of the tree-level ${\cal O}(\as)$ term 
alone, which is what we observe. We conclude this
discussion by noting that other definitions can be given of the
uncertainties associated with mass scales (e.g., variations may be
summed linearly rather than in quadrature), without this changing the
pattern found here. We also point out that the scale dependences
of the rapidity observables which we shall discuss below are rather 
featureless (i.e., scale variations are independent of rapidities), 
and therefore they will not be shown in what follows.

In fig.~\ref{fig:y} we consider the rapidity distributions
of the $W^+$ boson (left pane), and of the hardest jet of the event
(right pane). In the absence of any cuts in $\pt$, the boson rapidity
is an inclusive variable unaffected by large logarithms,
and both NLO computations and MC simulations should predict it
relatively well. We do indeed see an overall consistency among
\NLPY, \NLHW, and \PY\ standalone. Differences among the three 
predictions are larger in the case of the rapidity of the hardest jet,
since this is a less inclusive variable w.r.t. the $W$ rapidity,
and is also more sensitive to hadronization corrections.
The \PY\ standalone results have been rescaled by the NLO K-factor,
\mbox{$\sigma_{\sss NLO}/\sigma_{\sss LO}$}, before any cuts are
applied and jets are reconstructed. 

Finally, in fig.~\ref{fig:Dely} we consider the difference between
the rapidities of the $W^+$ boson and of the hardest jet, for two different
cuts on the transverse momentum of the hardest jet. The \PY\ standalone
results have been rescaled in the same way as in fig.~\ref{fig:y}. This 
observable and its analogues (with $y_W$ replaced by $y_S$, $S$ being the 
system emerging from the hard process at the Born level - e.g. the $t\bar{t}$
pair in top-pair production) has attracted some attention in the 
past (see e.g. ref.~\cite{Alioli:2008gx} for a recent discussion),
owing to the fact that \NLHW\ has a different behaviour around
$y_S-y_j\simeq 0$ w.r.t. the underlying NLO computation -- the 
former being flatter than the latter, or having a dip, depending
on the nature of the system $S$. The ``flatness'' or the presence
of a dip is a feature of MC simulations, and more specifically
is a consequence of the choices made for initial conditions,
as was done here for \PY\ in eq.~(\ref{maxvirt}).
It has to be stressed that this is true for both \HW\ and \PY, as 
the dashed histogram on the right pane of fig.~\ref{fig:Dely} shows, 
and therefore has nothing to do with the presence of dead 
zones\footnote{Negative weights in MC@NLO are also not an issue. 
This is {\em a fortiori} true in \NLPY, where they are basically 
negligible.} as such in \HW. In \PY, the choice of initial conditions
is much more flexible than in \HW, and one can make the dip of
fig.~\ref{fig:Dely} disappear by choosing a large-enough scale
as the maximum virtuality allowed for the shower. In doing so, however,
one may extend the collinear approximation, that is the core of
both \HW\ and \PY, outside its proper range of validity, and we
consider the choice made in eq.~(\ref{maxvirt}) a sensible one.
$\clH$ events in MC@NLO, or matrix element corrections in \HW\ and \PY, 
will give a substantial contribution to the region $y_S-y_j\simeq 0$ 
(see the solid and dotted histograms in fig.~\ref{fig:Dely}), where the
predictions will become closer (w.r.t. those of MC simulations without
matrix element corrections) to the pure-NLO results. Differences will
in general remain, that can be formally ascribed to effects beyond NLO.

\begin{figure}[htb]
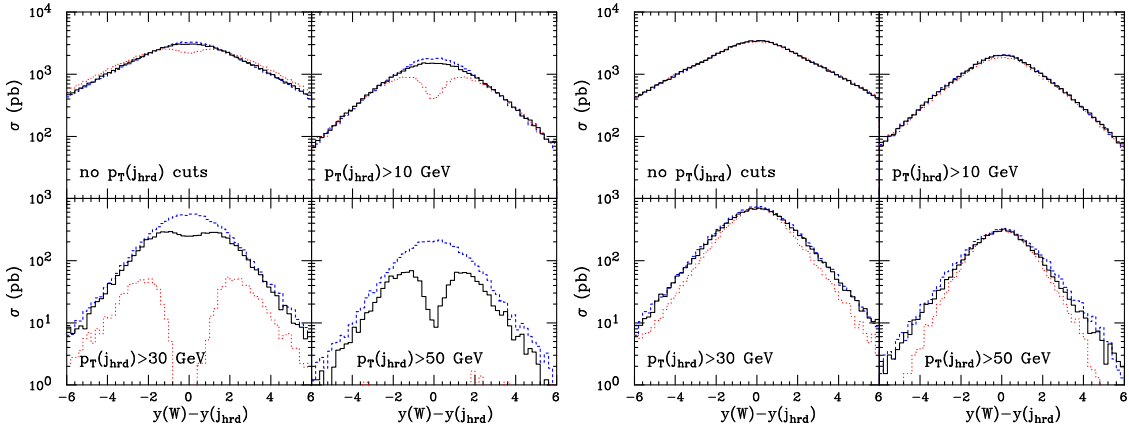

  \begin{center}
    \epsfig{figure=W-hj_pythia.eps,width=0.49\textwidth}
    \epsfig{figure=W-hj_mcatnlo.eps,width=0.49\textwidth}
\caption{\label{fig:Dely2} 
Rapidity differences between the $W$ and the hardest jet of the event,
with \PY\ (left pane) and \NLPY\ (right pane), for different values of
$f$ in eq.~(\ref{maxvirt2}). Four cuts on the $\pt$ of the jet are 
considered.}
  \end{center}
\end{figure}

In order to further this argument, we consider an extension of the choice
made in eq.~(\ref{maxvirt}). Namely, we parametrize the maximum virtuality
allowed in the shower in terms of a number $f$ introduced as follows:
\beq
\abs{t_i}\le \left(f\mV\right)^2\;\;\;\;\Longrightarrow\;\;\;\;
\stepf_\pm=\stepf\!\left(y-\frac{1-(1+2f^2)x}{1-x}\right).
\label{maxvirt2}
\eeq
The plots shown so far have therefore been obtained with $f=1$.
In fig.~\ref{fig:Dely2} we consider again the difference in rapidity
of fig.~\ref{fig:Dely}, with $f=1/2$ (dotted red histograms) and
with $f=2$ (dashed blue histograms), together with our default
choice $f=1$ (solid black histograms); we point out that for
the values of $f$ considered here the fraction of negative 
weights is basically a constant. Two additional $\pt$ cuts are
also considered on top of those of fig.~\ref{fig:Dely}.
The results obtained with \PY\ feature a very large sensitivity
to the choice of $f$, while those obtained with \NLPY\ are fairly 
stable. One may be tempted to use the results of fig.~\ref{fig:Dely2} 
in order to ``tune'' the parameter $f$, and obtain a \PY\ prediction
in decent agreement with that of \NLPY.
However, this is an {\em a posteriori} procedure, which can be justified
phenomenologically, but which implies that any possibility is given up
of a sensible estimate of the theoretical uncertainties affecting the
observable considered here; also, such a procedure is in general
observable-dependent. Indeed, the left pane of fig.~\ref{fig:Dely2} 
shows that, if no information on $2\to 2$ matrix elements is included,
this rapidity difference basically cannot be {\em predicted} by Monte 
Carlos at moderate and large $\pt$'s.
This is not surprising, given that one is attempting to use
the collinear approximation outside its range of validity: the correct
result can be recovered (which is equivalent to choosing the argument
of a logarithm so as its numerical value coincides with a given constant), 
but only in a heuristic way. The situation improves if the correct information 
on matrix elements is used, as in \NLPY. This is analogous to what happens 
when one varies the renormalization and factorization scales, 
at the LO and the NLO levels. As in the case of scale variations, extreme 
values for $f$ will lead to problems; however, $f={\cal O}(1)$ appears
to be a safe choice, allowing one to realistically estimate theoretical
uncertainties. We conclude by mentioning that other observables
(such as those shown in fig.~\ref{fig:pt} and~\ref{fig:y}) display
the same pattern of dependence on $f$ as the observable in
fig.~\ref{fig:Dely2}. We shall discuss this issue further~\cite{HiggsPY} 
by considering also the case of Higgs production, where the effects 
are more pronounced. 

\section{Conclusions\label{sec:concl}}
We have presented the construction of the matching between an NLO
QCD computation and the virtuality-ordered \PY\ Monte Carlo, according
to the MC@NLO formalism. We have limited ourselves to considering only 
the case of initial-state radiation, and applied the formalism to the
study of $W$ hadroproduction. Owing to the different structures of
\HW\ and \PY, the short-distance MC@NLO cross sections used
to generate hard events are different in the two cases. However, the
FKS subtraction, which is the method used in MC@NLO for dealing with
infrared singularities, needs no modifications, and is able to treat 
both MCs. Likewise, no changes are needed in the MC@NLO formalism,
and the matching with \PY\ is achieved by performing a process-independent
calculation.

For the process considered in this paper, the fraction of negative 
weights in \NLPY\ is much smaller than that in \NLHW. From the physical 
viewpoint, the pattern of the comparison between MC@NLO, pure-NLO, 
and PSMC results is the same for \NLPY\ as for \NLHW\ --- MC@NLO shows 
the same behaviour as the NLO or MC where either one is most reliable, 
with a smooth transition between the hard and soft-collinear emission regions.

Although the process studied in this paper is particularly simple, the
formulae given here will be basically sufficient for performing the
matching in the case of more complicated reactions (and limited to 
initial-state branchings only). Therefore, the results presented here
are the first step towards the construction of \NLPY\ for generic
processes, where also final-state radiation is present, and towards 
the extension of this formalism to the $p_T$-ordered version of \PY. 

\section*{Acknowledgments}
We would like to thank Torbj\"orn Sj\"ostrand, Peter Skands, and
Bryan~Webber for helpful discussions and comments on the manuscript.
The work of P.~T. is supported by the Swiss National Science Foundation.

\end{document}